# Quantifying Inactive Lithium in Lithium Metal Batteries

Chengcheng Fang[a, §], Jinxing Li[b, §], Minghao Zhang[b], Yihui Zhang[a], Fan Yang[c], Jungwoo Z. Lee[b], Min-Han Lee[a], Judith Alvarado[a,d], Marshall A. Schroeder[d], Yangyuchen Yang[a], Bingyu Lu[b], Nicholas Williams[c], Miguel Ceja[b], Li Yang[e], Mei Cai[e], Jing Gu[c], Kang Xu[d], Xuefeng Wang[b], and Ying Shirley Meng[a,b *]

a. Materials Science and Engineering Program, University of California San Diego, La Jolla, CA, 92093, USA.

b. Department of NanoEngineering, University of California San Diego, La Jolla, CA, 92093, USA.

c. Department of Chemistry and Biochemistry, San Diego State University, San Diego, CA, 92182, USA.

d. Electrochemistry Branch, Sensors and Electron Devices Directorate, U.S. Army Research Laboratory, Adelphi, MD, 20783, USA.

e. General Motors Research and Development Center, 30470 Harley Earl Blvd., Warren, MI 48092, USA

**Abstract:**

Inactive lithium (Li) formation is the immediate cause of capacity loss and catastrophic failure of Li metal batteries. However, the chemical component and the atomic level structure of inactive Li have rarely been studied due to the lack of effective diagnosis tools to accurately differentiate and quantify $Li^+$ in solid electrolyte interphase (SEI) components and the electrically isolated unreacted metallic $Li^0$, which together comprise the inactive Li. Here, by introducing a new analytical method, Titration Gas Chromatography (TGC), we can accurately quantify the contribution from metallic $Li^0$ to the total amount of inactive Li. We uncover that the $Li^0$, rather than the electrochemically formed SEI, dominates the inactive Li and capacity loss. Using cryogenic electron microscopies to further study the microstructure and nanostructure of inactive Li, we find that the $Li^0$ is surrounded by insulating SEI, losing the electronic conductive pathway to the bulk electrode. Coupling the measurements of the $Li^0$ global content to observations of its local atomic structure, we reveal the formation mechanism of inactive Li in different types of electrolytes, and identify the true underlying cause of low Coulombic efficiency in Li metal deposition and stripping. We ultimately propose strategies to enable the highly efficient Li deposition and stripping to enable Li metal anode for next generation high energy batteries.

**Main Text:**

To achieve the energy density of 500 Wh/kg or higher for next-generation battery technologies, Li metal is the ultimate anode, because it is the lightest metal on earth (0.534 g cm$^{-3}$), delivers ultra-high theoretical capacity (3860 mAh g$^{-1}$), and has the lowest negative electrochemical potential (-3.04 V *vs.* SHE)[1]. Yet, Li metal suffers from dendrite growth and low Coulombic efficiency (CE) which have prevented the extensive adoption of Li metal batteries (LMBs)[2–4]. Since the first demonstration of a Li metal battery in 1976[5], tremendous effort has been made in preventing dendritic Li growth and improving CE, including electrolyte engineering[6–9], interface protection[10] and substrate architecture[11]. While dense Li can be achieved without any dendrites during the plating process, the stripping process will eventually dominate the CE thus the reversibility of Li metal anode.

The formation of inactive Li, also known as "dead" Li, is the immediate cause of low CE, short cycle life and violent safety hazard of LMBs. It consists of both (electro)chemically formed Li$^+$ compounds in the solid electrolyte interphase (SEI) and the electrically isolated unreacted metallic $Li^0$[12,13]. It is generally assumed that the low CE is dominated by the continuous repairing of SEI fracture that consumes both electrolyte and active Li metal[14]. However, the actual contribution of capacity loss from the SEI formation has never been quantified. Consequently, efforts may be misdirected as we search for solutions to the low CE. Differentiating and quantifying the $Li^+$ and $Li^0$ remaining on the electrode after stripping, therefore, becomes the key to understanding the mechanisms leading to capacity decay; which is challenging due to the lack of proper characterization tools. Microscopy and other imaging tools, such as operando optical microscopy[15], *in-situ* environmental transmission electron microscopy (TEM)[16,17], X-ray microtomography[18] and magnetic resonance imaging (MRI)[19], have been used to visualize the dynamic growth of Li dendrites, but most of them only provide morphological perspective with little chemical information. Nuclear magnetic resonance (NMR) utilized as a quantitative and operando tool to reveal the Li microstructure formation with a resolution of tens of micrograms of Li, however the "skin-depth" problem based working mechanism makes it only effective to quantify dendritic/mossy Li[20]. The X-ray photoelectron spectroscopy (XPS)[21] and cryogenic TEM[22,23] have the capability to distinguish between the $Li^+$ in SEI components and metallic $Li^0$, but are limited to surfaces or local regions and cannot be used for global quantitative measurements. It is thus of essential significance to develop a set of experimental tools that both qualitatively profile the micro/nano-structures of inactive Li, and quantitatively differentiate the SEI $Li^+$ and metallic $Li^0$.

Here we combine multiscale characterization techniques to achieve this goal. A new analytic method, Titration Gas Chromatography (TGC), is demonstrated to accurately determine the quantity of isolated metallic $Li^0$ content in inactive Li down to 1 microgram (µg). Advanced cryogenic FIB-SEM and TEM are used to probe the microstructure and nanostructure of inactive Li in both SEI components and the isolated metallic form, providing crucial complimentary information on nanoscale. Combining these results, we propose the formation mechanism of inactive Li and strategies to mitigate it and thus to achieve high CE.

**TGC for Quantitative Analysis**

Inactive Li is believed to contain diverse $Li^+$ compounds within the SEI, such as LiF, $Li_2CO_3$, $Li_2O$, $RCO_2Li$[24,25], and remaining metallic $Li^0$ which is isolated by SEI from the electronic conductive pathway. The key difference between the SEI $Li^+$ and metallic $Li^0$ that we take advantage of is their reactivity, where only the metallic $Li^0$ reacts with protic solvents (e.g. $H_2O$) and generate hydrogen gas ($H_2$). Therefore, we combine $H_2O$ titration and gas chromatography herein referred to as TGC (schematic process in Fig. 1), which is able to quantify the content of metallic $Li^0$ based on the following reaction:

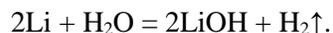

$$2Li + H_2O = 2LiOH + H_2\uparrow.$$

Coupling with an advanced $H_2$ detector, this method can accurately determine a minimum $H_2$ concentration of 100 ppm, corresponding to 1 µg of metallic $Li^0$ in the designed system. Complete methodologies are illustrated in the Supplementary Information.

We first validate the method by using commercial Li metal with known mass. The result shows that Li metal mass is linear to ($R^2 = 99.8\%$) the detected $H_2$ amount (Fig. S1), indicating the feasibility and accuracy of the TGC system.

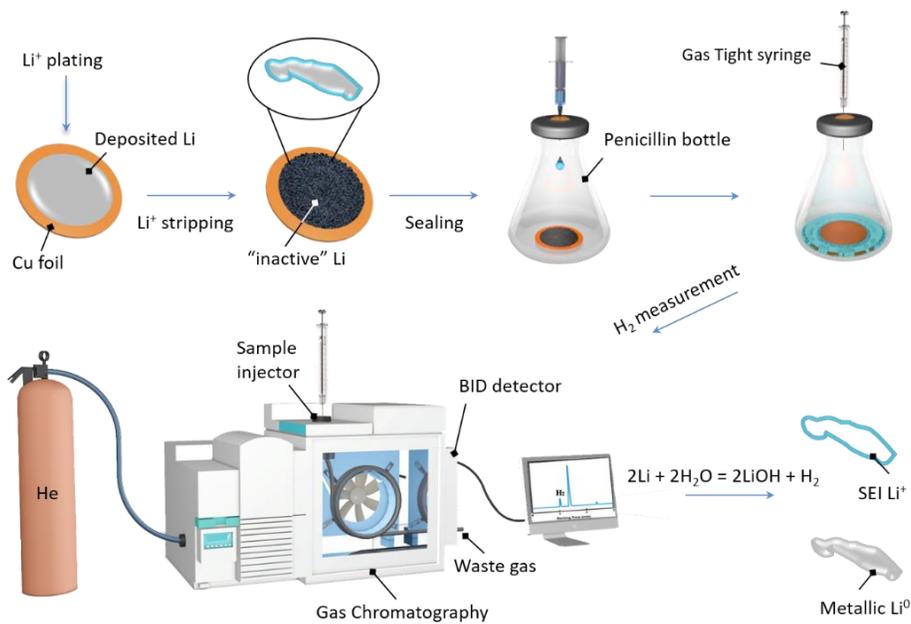

**Fig. 1. Schematic of the working principle of the TGC method.** Combing $H_2O$ titration on inactive Li sample and $H_2$ quantification by GC, metallic $Li^0$ amount is determined based on the chemical reaction $2Li + 2H_2O = 2LiOH + H_2$.

We then applied TGC to correlate the origin of inactive Li with the CE in Li||Cu half cells. Noting that the CE of Li metal is profoundly influenced by the electrolyte properties and current density, two representative electrolytes, the high-concentration electrolyte (HCE, 4.6m LiFSI + 2.3m LiTFSI in DME) and commercial carbonate electrolyte (CCE, 1M $LiPF_6$ in EC/EMC), were compared at three stripping rates (0.5, 2.5 and 5.0 mA $cm^{-2}$, all plating was done at 0.5 mA $cm^{-2}$ for 1 mAh $cm^{-2}$). Fig. 2a and b show their average CE and representative voltage profiles during the first cycle. The small error bars indicate good consistency among the cells. As expected, the HCE exhibits higher CE and better rate performance than the CCE, which is consistent with previous reports [8,26]. The total amount of inactive Li is equal to the capacity loss between the plating and stripping processes, displaying a liner relationship with CE in Fig. 2d. The content of the metallic $Li^0$ was directly measured by the TGC method. As summarized in Fig. 2c, in the HCE, the capacity loss from inactive metallic $Li^0$ is about 60% at different stripping rates. This is consistent with the similar CE at various stripping rates. Whereas in the CCE, it contributes over 90%, especially at high stripping rate. Surprisingly, the amount of unreacted metallic $Li^0$ exhibits a linear relationship with loss of CE, as shown in Fig. 2e, which is almost independent from the testing conditions. This implies that the CE loss is governed by the formation of inactive metallic $Li^0$. Thus, the SEI $Li^+$ amount (Fig. 2f) deducted from above remains low and relatively constant under various stripping rates. The TGC measurement unequivocally indicates that the inactive $Li^0$ dominates the inactive Li and the capacity loss.

Further examining the SEI components by XPS, we found that stripping rates have negligible impact on the relative contributions from SEI components and contents, as specified in Fig. S7. The TGC quantification analysis and XPS results clearly indicate that the contribution from the SEI to the global content of inactive Li is not as significant as commonly believed in previous studies[7,27,28].

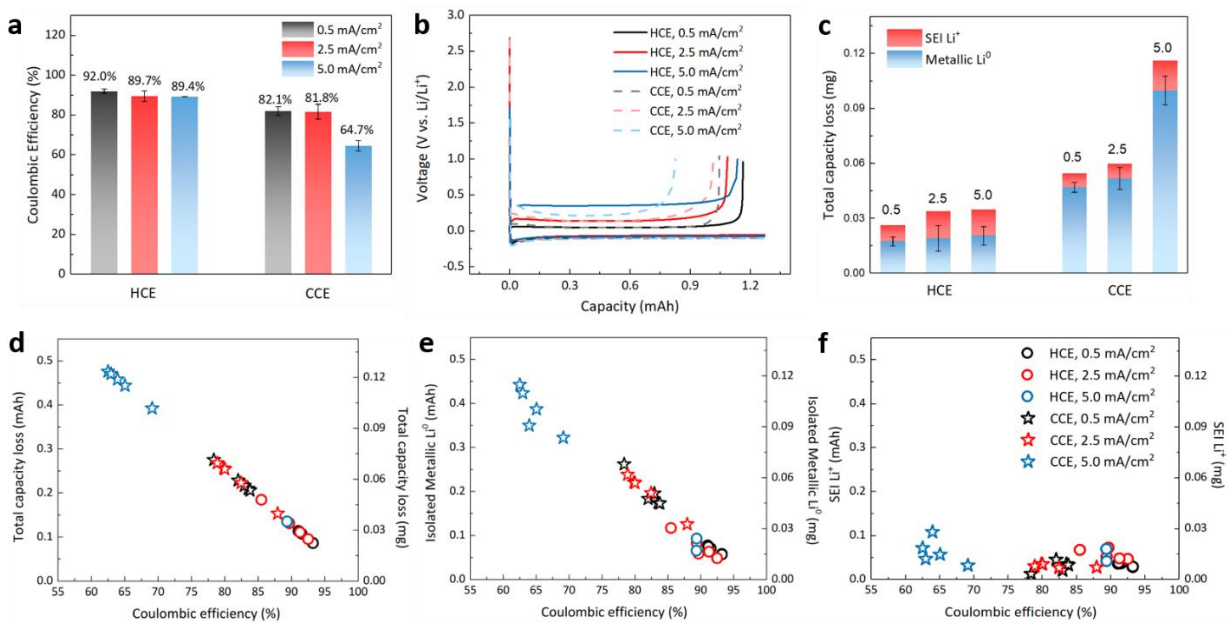

**Fig. 2 Quantitative differentiation of inactive Li by the TGC method.** (**a**) Average CE of Li‖Cu cells under different testing conditions. (**b**) Representative voltage profiles of Li‖Cu cells under different testing conditions in the first cycle. (**c**) Summarized quantitative contribution of capacity loss from the SEI Li$^+$ and metallic Li$^0$. (**d**) Total capacity loss as a function of CE. (**e**) Inactive metallic Li amount measured by the TGC method as a function of CE. (**f**) Calculated SEI Li$^+$ amount as a function of CE.

In order to elucidate the formation mechanism of inactive Li, cryogenic FIB-SEM and TEM are combined to explore the micro- and nano-structures of inactive Li. Cryogenic protection is critical here because the highly reactive Li metal is not only sensitive to the electron beam, but also is apt to react with the FIB incident Ga ion beam to form a Li$_x$Ga$_y$ alloy at room temperature[29]. Completely different morphologies were observed at the various stripping rates even though the same morphology is formed during plating at the same rate (Fig. S5a-c). As the stripping rate increases, the morphology of inactive Li in the HCE evolves from thin, dense sheets (Fig. 3a) to local clusters (Fig. 3c) with a thickness increased from 500 nm (Fig. 3d) to 2 μm (Fig. 3f). For the CCE, a mossy-like morphology with interwoven ribbons remain after stripping (Fig. 3g-i), but becomes thicker at higher stripping rates. It is worth noting that these residues also have poor contact with the current collector, indicating the loss of a direct electronically conducting pathway.

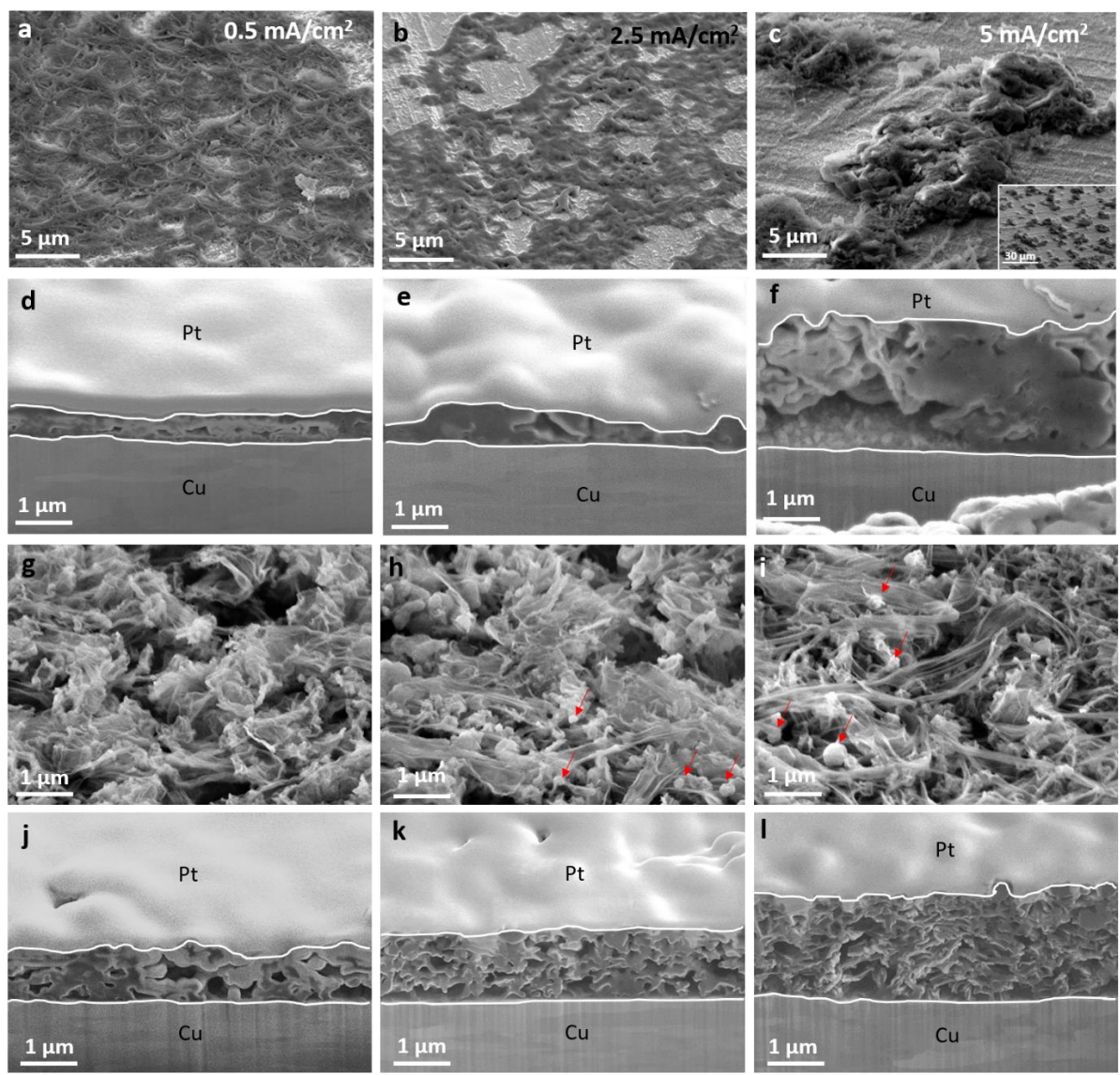

**Fig. 3 Microstructures of inactive Li in the HCE (a-f) and CCE (g-l) by Cryo FIB-SEM.** (**a-c**) and (**g-i**) are top view of the inactive Li at 52º tilted stage. (**d-f**) and (**j-l**) are cross-sections. Each column represents a stripping rate. (**a**, **d**, **g** and **j**) are at 0.5 mA cm$^{-2}$. (**b**, **e**, **h** and **k**) are at 2.5 mA cm$^{-2}$. (**c**, **f**, **i** and **l**) are at 5.0 mA cm$^{-2}$.

Cryo-TEM has recently been demonstrated to be a powerful technique to probe the nanostructure of Li metal as well as the SEI[22,23,30]. The cryogenic protection minimizes the beam damage to the brittle Li metal while preserve its intrinsic properties. We further used cryo-TEM to investigate the nanostructure of the inactive Li in both electrolytes after stripping at 0.5 mA cm$^{-2}$. The low magnification images show that the inactive Li in the HCE has a sheet-like morphology (Fig. 4a) while that in the CCE is ribbon-like (Fig. 4f), consistent with the above FIB-SEM observations. Some parts of the ribbons were thinner with low contrast, indicating the removal of bulk metallic Li$^0$, while other parts of the ribbons maintain similar size and contrast to the deposited Li ribbon (Fig. S6), suggesting the trapping of isolated Li$^0$ within SEI matrix. Images with increased magnification allow the differentiation of Li$^0$ and SEI. In the HCE, a very small area of crystalline metallic Li$^0$ is embedded in many SEI layers and it occupies only about 1% area of the inactive

Li in Fig. 4b. This suggests that most of deposited Li metal has been successfully stripped, corresponding to the higher CE of the HCE. Fig. 5d further demonstrates that the $Li^0$ has a size of about 5 nm and is isolated by the SEI. The SEI components were determined by matching the lattice spacing in TEM images with their fast Fourier transform (FFT) patterns, which are dominated by the crystalline $Li_2O$ and LiF, as well as other amorphous organic species as indicated by XPS analysis (Fig. S7). In sharp contrast, the CCE leads to a much larger area (about 30%) of crystalline $Li^0$ wrapped by the SEI (Fig. 4g and 4i). Its lattice spacing is measured to be 0.246 nm in Fig. 4i, which agrees with the (110) distance of *bcc* Li and confirms the presence of the $Li^0$. The FFT patterns in Fig. 4g indicate that the dominant crystalline SEI component in the CCE is $Li_2O$. The Cyro-TEM observation confirms that the metallic $Li^0$ in inactive Li is indeed wrapped by insulating SEI layers becoming inactive. Even though the SEI dominates the observed areas in the TEM due to its high surface area, it is the metallic $Li^0$ that constitutes the primary weight content of the inactive Li as indicated by the TGC results.

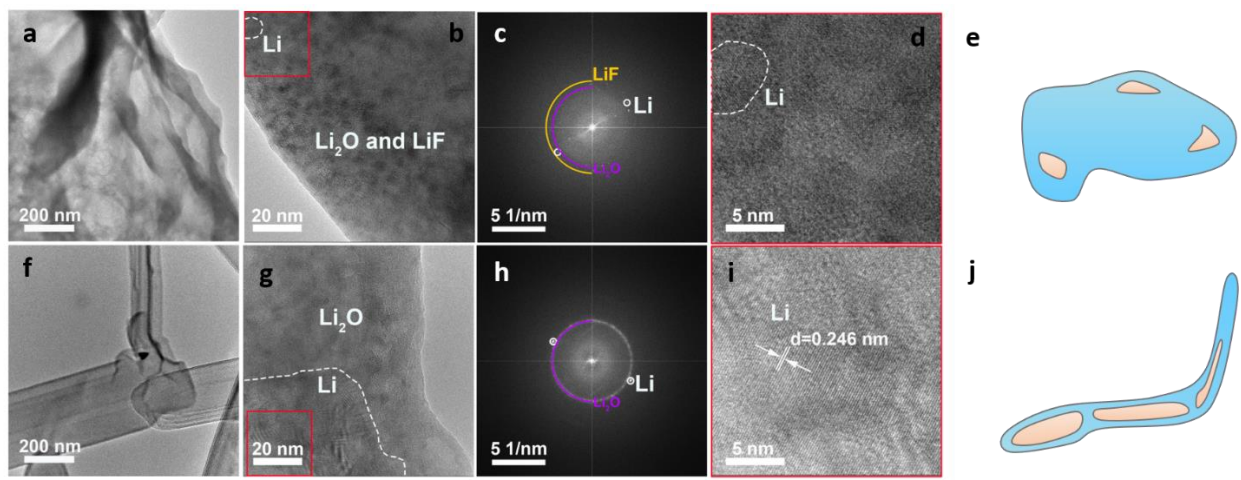

**Fig. 4 Nanostructures of inactive Li generated in HCE (a-e) and CCE (f-j) electrolytes at 0.5 mA cm$^{-2}$ by Cryo-TEM.** **(a)** In the HCE electrolyte, sheet-like inactive Li is generated. **(b)** A small volume of metallic Li is wrapped by SEI. **(c)** FFT patterns indicate the SEI contains crystalline LiF and $Li_2O$. **(d)** High-resolution TEM image shows the metallic Li is encapsulated by SEI, which consists of crystalline species embedded in an amorphous matrix. **(e)** Schematic of inactive Li nanostructure in the HCE electrolyte. Small area of metallic $Li^0$ is embedded in a sheet-like SEI layer. **(f)** The inactive Li generated in the CCE electrolyte has a ribbon-like morphology. **(g)** Inactive metallic $Li^0$ is wrapped by SEI, with a substantially larger amount than that in the HCE electrolyte. **(h)** FFT patterns indicate the crystalline components in SEI formed in the CCE electrolyte is mainly $Li_2O$. **(i)** High-resolution TEM image shows the lattice fringes of crystalline metallic Li. **(j)** Schematic of the inactive Li nanostructure in the CCE electrolyte. A large bulk of metallic $Li^0$ is wrapped in ribbon-like SEI layer.

**Discussion**

A combination of state-of-the-art characterization methods, TGC quantification, XPS, Cryo-FIB-SEM and Cryo-TEM, demonstrates a comprehensive picture of inactive Li across multiple scales. We discover that inactive Li is mainly comprised of unreacted $Li^0$ rather than the $Li^+$ in SEI, the contribution of which is overstated in the previous research. This misunderstanding is likely caused by the high surface area of SEI that would be relatively easier to detect by prior analytic methods. The titration method here unequivocally indicates that the amount of isolated but unreacted $Li^0$ dominates the weight percentage of the inactive Li, as evidenced by its linear correlation with the loss of CE. Cryo-TEM results further verify that the $Li^0$ is entrapped in the insulating SEI matrix, isolated from the conductive network of the bulk electrode, and thus become inactive. Overall, these metallic $Li^0$ are responsible for the majority of the loss of capacity and CE.

Correlating the inactive metallic $Li^0$ content with the micro and nanostructure of the inactive Li formed under different conditions, we propose a model for the formation mechanism of the inactive Li as well as the stripping mechanism of Li metal. Two processes are involved in the stripping: 1) $Li^+$ dissolution. Under the electric field, metallic $Li^0$ is oxidized to $Li^+$, which diffuses through the SEI layers and dissolves into the electrolyte. 2) SEI collapse. When the Li is removed, the SEI simultaneously shrinks and collapses towards the current collector. During these two dynamic processes, we emphasize an ignored but crucial aspect, the structural connection, which is defined as the capability of the active Li to maintain an electronic conductive network. Our Cryo-FIB-SEM and Cryo-TEM images show that inactive $Li^0$ either directly disconnected from the current collector or entrapped by the insulating SEI will lead to the loss of structural connection, which in turn is determined by both the micro and nanostructure of the deposited Li. Obviously, for a Li deposit with whisker morphology and large tortuosity (CCE case) as shown in Fig. 5a, both manners of losing structural connection can easily occur due to the undesired microstructure, resulting in poor structural connection and more unreacted $Li^0$ trapped in SEI during the stripping process. In contrast, the dense Li with chunky morphology and low tortuosity (HCE case), as exhibited in Fig. 5b, has bulk integrity to maintain a structural connection and an intimate contact with the current collector, resulting in reduced amount of isolated $Li^0$ and high CE. This is further evidenced by an advanced novel electrolyte with columnar microstructure and minimum tortuosity. Fig. S8 shows a cross-section morphology of Li deposits generated in the advanced electrolyte, which could deliver a first cycle CE as high as 96.2%. Based on the proposed model, we ascribe the excellent performance to the columnar microstructure with minimum tortuosity, which significantly enhances the structural connection.

Maintaining a good structural connection is the key to reducing the amount of the inactive metallic $Li^0$ and increasing the CE, which can be realized by controlling the micro and nanostructure of the plated Li deposits. Moreover, the structural connection can be further facilitated by applying external pressure. It is reported that applying slight stacking pressure helps improve cycling performance significantly[31]. In our proposed model, pressure could promote the structure collapse towards the current collector and thus maintain a good structural connection, mitigating the generation of inactive metallic $Li^0$.

Based on the above observation and discussion, we propose the following strategies that could improve CE. First, an ideal architecture of deposited Li (Fig.5c) would promote a good structural connection and mitigate the inactive Li formation, especially the formation of inactive $Li^0$. 1) The Li deposits should retain a columnar microstructure with a large granular size and minimum tortuosity. 2) The SEI should be homogeneous in both component and distribution for uniform $Li^+$ dissolution. In addition to the microstructure, applying proper external pressure helps to keep a good structural connection. Fast stripping rate could accelerate the $Li^+$ dissolution but may be harmful to structural connection due to the fast dynamic. Three-dimensional (3D) hosts that maintain electronic pathway and low tortuosity could potentially contribute to initiating a good structural connection and guiding the Li plating and stripping.

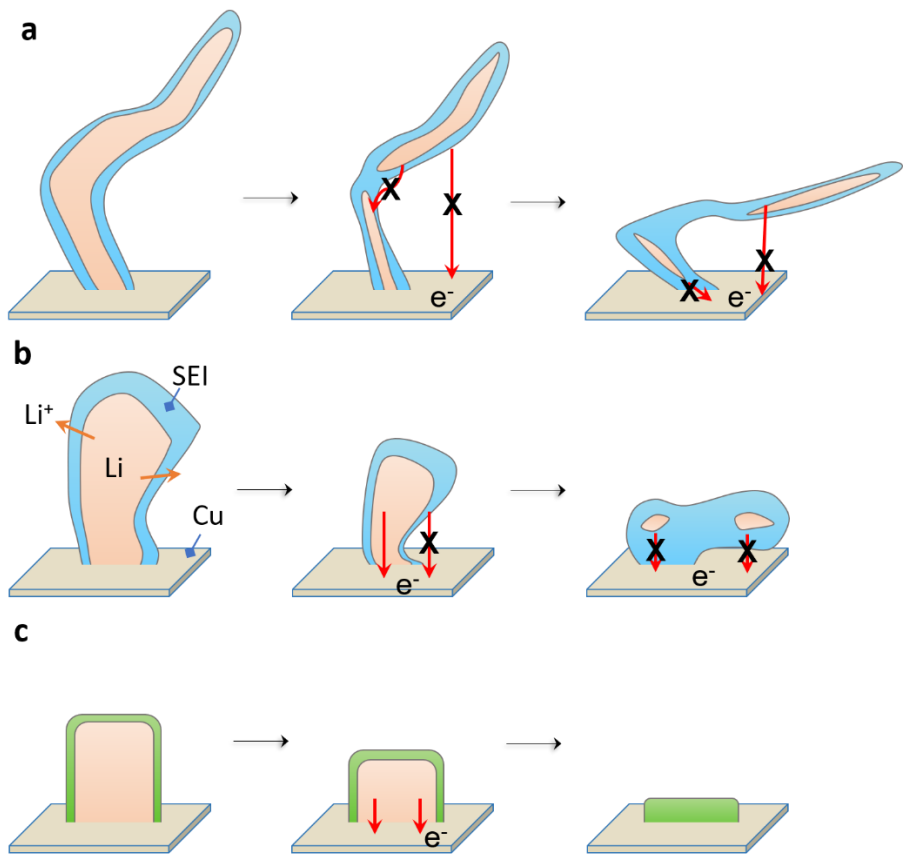

**Fig. 5 Schematic of inactive Li formation mechanism in different electrolytes based on TGC quantification, Cryo-FIB and Cryo-TEM observation.** (**a**) Li deposits with whisker morphology and large tortuosity are more likely to lose electronic connection and maintain poor structural connection, leaving large amount of unreacted metallic Li trapped in SEI. (**b**) Li deposits with large granular size and less tortuosity tend to maintain a good structural electronic connection, only bits of metallic Li are stuck in tortuous SEI edges. (**c**) An ideal Li deposit should have a columnar microstructure with a large granular size and minimum tortuosity and homogeneous distribution of SEI component, facilitating a complete dissolution of active $Li^0$.

**Conclusion**

In summary, this work presents a novel quantitative methodology to accurately differentiate and quantify the inactive Li on multi-scales. The TGC results reveal that the unreacted $Li^0$ is mainly responsible for the capacity loss of LMBs rather than $Li^+$ in SEI components, contrary to previous beliefs. The morphological study by Cryo-FIB-SEM shows that the microstructures, such as particle sizes and tortuosity, of the deposited Li play key roles in maintaining the structural electronic connection. Cryo-TEM confirmed that the isolated particles of $Li^0$ is indeed trapped in the electronically insulating SEI. Correlating these observations, we proposed the formation mechanism of inactive Li, revealing that the true underlying cause of capacity loss in LMBs is due to large amount of metallic $Li^0$ becoming trapped in SEI with tortuous microstructures. We offered strategies to mitigate inactive Li formation and thus significantly improve CE. With an ideal columnar microstructure, it could be possible that the capacity loss will only come from the minor SEI formation in the initial cycle and the anode-free battery can thus be realized. The versatile characterization tools proposed here can be further extended to investigate inactive Li properties under different conditions, such as after long term cycling, with different electrolytes and various extreme temperature conditions, serving as a standard methodology to evaluate the strategies that improve the performance of Li metal. We also expect that our methods will be expanded for systematic study of other metal such as sodium, magnesium and zinc metal batteries with unprecedented achievements.

**Acknowledgements**

This work was supported by the Office of Vehicle Technologies of the U.S. Department of Energy through the Advanced Battery Materials Research (BMR) Program (Battery500 Consortium) under Contract DE-EE0007764. Cryo-FIB was performed at the San Diego Nanotechnology Infrastructure (SDNI), a member of the National Nanotechnology Coordinated Infrastructure, which is supported by the National Science Foundation (grant ECCS-1542148). We acknowledge the UC Irvine Materials Research Institute (IMRI) for the use of the Cryo-Electron Microscopy and XPS facilities, funded in part by the National Science Foundation Major Research Instrumentation Program under Grant CHE-1338173. C.F. thanks Mr. Daniel M. Davis for his valuable suggestions on the manuscript.

**Authors Contributions**

C.F., J.L., X.W. and Y.S.M. conceived the ideas. C.F. designed and implemented the TGC system. C.F. designed and performed the TGC, cryo-FIB-SEM, XPS experiments and data analysis. C.F., Y.Z. and M.C. prepared samples for characterizations. J.Z.L. and Y.Y. helped set up cryo-FIB instrumentation. F.Y., N.W. and J.G. helped with GC set up and calibration. C.F. and M.L. performed the RGA experiment. J.A., M.A.S. and K.X. formulated and provided the HCE electrolyte. L.Y. and M.C. formulated and provided the GM electrolyte. J.L. and C.F. wrote the manuscript. All authors discussed the results and commented on the manuscript. All authors have given approval to the final version of the manuscript.